\documentclass[a4paper]{article}

\usepackage[english]{babel}
\usepackage[utf8]{inputenc}
\usepackage{fullpage}
\usepackage{amsmath,amsthm,amsfonts,amssymb,amscd}
\usepackage{graphicx}
\usepackage{pst-all}
\usepackage[colorinlistoftodos]{todonotes}
\usepackage{setspace}
\usepackage{pstricks}
\usepackage{bbold}
\usepackage{caption}
\usepackage{subcaption}
\usepackage{pict2e}
\usepackage{lastpage}
\usepackage{enumerate}
\usepackage{fancyhdr}
\usepackage{mathrsfs}
\usepackage{xcolor}
\usepackage{graphicx}
\usepackage{listings}
\usepackage{hyperref}
\usepackage{amsthm}
\usepackage{bbm}
\usepackage{xfrac}

\theoremstyle{definition}
\newtheorem{definition}{Definition}[section]

\title{Distributed Verifiers in PCP}

\author{Nagaganesh Jaladanki and Wilson Wu}

\DeclareMathOperator{\Had}{Had}

\newcommand{\PCP}{\mathsf{PCP}}
\newcommand{\IP}{\mathsf{IP}}

\newcommand{\LCP}{\mathsf{LCP}}
\newcommand{\logLCP}{\mathsf{logLCP}}
\newcommand{\dIP}{\mathsf{dIP}}

\newcommand{\dMAM}{\mathsf{dMAM}}
\newcommand{\dPCP}{\mathsf{dPCP}}

\newcommand{\Sym}{\textsc{Sym}}
\newcommand{\Span}{\textsc{Span}}
\newcommand{\Nonbipartite}{\textsc{Nonbiparite}}
\newcommand{\Leader}{\textsc{Leader}}
\newcommand{\fm}{\mathcal{F}}
\newcommand{\lto}{\longrightarrow}
\newcommand{\trans}{^\top}

\begin{document}
\maketitle

\onehalfspacing

\begin{abstract}

Traditional proof systems involve a resource-bounded verifier communicating with a powerful (but untrusted) prover. Distributed verifier proof systems are a new family of proof models that involve a network of verifier nodes communicating with a single independent prover that has access to the complete network structure of the verifiers. The prover is tasked with convincing all verifiers of some global property of the network graph. In addition, each individual verifier may be given some input string they will be required to verify during the course of computation. Verifier nodes are allowed to exchange messaged with nodes a constant distance away, and accept / reject the input after some computation.

Because individual nodes are limited to a local view, communication with the prover is potentially necessary to prove \textit{global} properties about the network graph of nodes, which only the prover has access to. In this system of models, the entire model accepts the input if and only if every individual node has accepted.

There are three models in the distributed verifier proof system family: $\LCP$, $\dIP$, and our proposed $\dPCP$, with the fundamental difference between these coming from the type of communication established between the verifiers and the prover. In this paper, we will first go over the past work in the $\LCP$ and $\dIP$ space before showing properties and proofs in our $\dPCP$ system.

\end{abstract}

\section{Preliminaries}
For a graph $G$, we denote by $V(G)$ and $E(G)$ the vertex and edge sets of the graph, respectively. For some vertex $i\in V(G)$, we define the neighborhood $N(i)$ of $i$ as those vertices in $G$ adjacent to $i$, including $i$ itself. That is, $N(i)=\{j\in V(G)\mid j=i\lor (i,j)\in E(G)\}$.

In general, problem instances for each model will take the form of $(G,x)$ for $G$ in some family of graphs $\fm$ and $x:V(G)\lto \{0,1\}^*$ a function from vertices to binary strings. Intuitively, $\fm$ is a ``promise'' that $G$ has some structure, e.g. that it is connected, while $x$ is an input string for each verifier node. For a vertex subset $S\in V(G)$, we denote by $x|_{S}$ the restriction of $x$ to $S$.

For each proof system we discuss, the distributed verifier consists of $V_i$ for each vertex $i\in G$. Each verifier is \textit{local}, in that it makes a decision based only on the structure, inputs, and proofs of its neighborhood. That is, each verifier is a function of the format $V^{\pi_{N(i)}}_i(N(i), x|_{N(i)})$ where $\pi_{N(i)}$ is, abstractly, an oracle with access to the proofs sent to $N(i)$. The format of these proofs will be specified in more detail for each model we discuss.

We will consider several natural graph properties and languages, which we now define.
\begin{definition}
A graph $G$ is in $\Nonbipartite$ if it is not bipartite. That is, there exists no $2$-coloring of $G$.
\end{definition}

\begin{definition}
A graph $G$ is in $\Sym$ if some nontrivial automorphism exists on $G$, i.e. some nonidentity permutation $\pi:V(G)\lto V(G)$ exists such that $\pi$ is compatible with the structure of $G$, in that $\forall i,j\in V(G):(i,j)\in E(G)\longleftrightarrow (\pi(i),\pi(j))\in E(G)$.
\end{definition}

\begin{definition}
$\Leader$ is the language of graphs with a unique distinguished ``leader''. More explicitly, a graph with input strings $(G,x)$, with $x$ a function $x:V(G)\lto\{0,1\}$, is in $\Leader$ if $x(i)=1$ for exactly one vertex $i\in V(G)$, and $x(j)=0$ for all other vertices $j\neq i$.
\end{definition}

\begin{definition}
A graph and input $(G,x)$ is in $\Span$ if $x$ defines a valid spanning tree on $G$. That is, for each $i\in V(G)$ the string $x(i)$ identifies either a neighbor of $i$, supposedly its parent in the spanning tree, or with some unique string specifies that $i$ is the root of $T$. If the directed graph $T$ defined by $x$ is indeed a spanning tree on $G$, then $(G,x)\in\Span$.
\end{definition}

\section{Locally Checkable Proofs}

In the Locally Checkable Proofs ($\LCP$) model, the prover can only send a single different proof string to each verifier, after which there is no further communication with prover. Once this proof string is received, verifier nodes can communicate with their local neighborhood before accepting or rejecting the input.

Formally, the Prover is a function $P: V(G) \lto \{0, 1\}^*$ that associates every vertex to the proof string that that the Prover sends. The verifier $\mathcal{A}$ is a computable function $V^{\pi_{N(i)}}_i(N(i), x|_{N(i)})$ that has an oracle to the proofs and input strings of its local neighborhood. Note that the $\LCP$ model does not allow verifiers to use randomness to accept or reject the Prover's proof string. As a result, the following definition encapsulates the $\LCP$ model.

\begin{definition}
We say a given graph property $\mathcal{P} \subseteq \mathcal{F}$ admits locally checkable proofs if the two properties hold.
\begin{itemize}
    \item If $G \in \mathcal{P}$, then there exists a proof $P: V(G) \lto \{0,1\}^*$ such that all verifiers accept.
    \item If $G \notin \mathcal{P}$ then any proof $P: V(G) \lto \{0,1\}^*$ will have at least one verifier reject.
\end{itemize}
\end{definition}

Of particular interest in this model is the communication complexity between the prover and each verifier, which allows a complexity hierarchy to be defined within this model.

\begin{definition}
We define the class $\LCP(f)$ to consist of graph properties that admit a locally checkable proof where the communication between each verifier and the prover is up to $f(n)$ bits.
\end{definition}

This following levels in the complexity hierarchy are of interest: $\LCP(0), \LCP(1), \LCP(O(\log n))$, and $\LCP(O(\textsf{poly}(n))$, which were first described in a by a 2011 paper by G{\"o}{\"o}s and Suomela \cite{lcp}. We define $\logLCP$ as an alias for $\LCP(O(\log n))$. 

Intuitively, $\logLCP$ is an interesting complexity class as it allows the prover to send a concise yet non-negligible amount of information to each verifier to prove some global property about the graph.

\subsection{$\logLCP$ lower bounds}

An interesting property about the $\logLCP$ class is the existence of nontrivial lower bounds proved on proof size. Uniquely identifying every node in a graph of $n$ vertices takes a lower bound of $\log n$ bits, so one can potentially characterize the $\logLCP$ class as one that sends the identifiers of a constant number of nodes to each vertex in the proof string.

Several problems in the $\logLCP$ class, such as $\Span$, $\Nonbipartite$, or $\Leader$, have $O(\log n)$ bits of communication as a lower bound for proof strings.

The proof sketch for this is as follows. We can take several small cycles that form \textit{yes}-instances of a particular graph property and ``glue" them together to form a longer cycle that does not match the relevant graph property. The smaller \textit{yes}-instances, which each require fewer bits to convey node identifiers and proof labels, still hold true locally when glued together, causing all nodes to accept based off of their local neighborhood. This will lead to an acceptance, even though the glued longer cycle is not part of the language.

\section{Distributed Interactive Proofs}

The $\dIP$ model extends upon the $\LCP$ model by introducing the notion of interaction and randomness between each verifier and the prover. Every verifier and the prover are allowed to communicate by exchanging communication in a series of rounds. These rounds of communication may be interspersed by communication with other nodes in the local neighborhood of the verifier. Because of the introduction of randomness, the prover-verifier system is not guaranteed to always produce the right answer such as in the $\LCP$ model.

\begin{definition}
We say a given graph property $\mathcal{P} \subseteq \mathcal{F}$ admits a distributed interactive proof if the two properties hold.
\begin{itemize}
    \item If $G \in \mathcal{P}$, then there exists a prover $P$ such that all nodes accept with probability greater than $\frac{2}{3}$.
    \item If $G \notin \mathcal{P}$ then for any prover, the probability that all nodes accept is less than $\frac{1}{3}$. 
\end{itemize}
\end{definition}

Ideas from traditional $\IP$ models can be extended to the $\dIP$ model as well. In particular, the paper proposing the $\dIP$ model worked extensively in the public-coin variant of the model, in which the verifiers share all generated randomness with the prover.\cite{dip_proposal} A prominent system used was the $\dMAM$ model. Communication complexity is denoted in this model as $\dMAM[f]$, where the number of bits transferred between the prover and verifier is upper bounded by $f(n)$. 

Intuitively, randomness and interaction seem to give this model more power, which may allow the net communication complexity to go down. Indeed, this is true, with results showing that $\Sym \in \dMAM[O(\log n)]$. This is direct improvement from the $\LCP$ model, which showed that $\Sym \notin \LCP(o(n^2))$.

\subsection{RAM Compiler}

The RAM compiler, introduced by Naor, Parter, and Yogev, is a general way of transforming traditional $\IP$ graph protocols with a centralized verifier into those that can be accepted by the network graph of verifiers \cite{ram_verifier}. 

At a high level, this reduction involves using the network graph as a RAM machine, with each individual verifier responsible for a small portion of the computation. The computation is checked to be globally correct with a specific reduction to Set Equality, for which a protocol is given in the paper. 

\section{Distributed PCP}
In the distributed $\PCP$ model we propose, the prover provides one global proof string which can be queried by each verifier in the graph. Intuitively, this model gains some power from the fact that the prover is forced to commit to a single, shared proof for all verifiers -- each verifier can both check that the proof is locally consistent with its neighborhood, and that it honestly encodes some global structure of the graph and vertex inputs. This sort of verifier strategy does not appear in $\LCP$ or $\dIP$, as in these classes the prover is free to send inconsistent proofs to different verifiers.

Formally, we define distributed $\PCP$ as follows:

\begin{definition}
Given a family of graphs $\mathcal{F}$ and graph language $L\subseteq\fm\times\{0,1\}^*$, we have $L\in\dPCP_{c,s}[l,r, q]$ if
\begin{itemize}
    \item \textbf{Completeness:} For any $(G,x)\in L$, there exists a proof $\pi$ with $|\pi|=l$ such that \\ $\Pr\left(\forall i\in V(G):V_i^{\pi}(x_{N(i)}=1\right)>c$.
    \item \textbf{Soundness:} For any $(G,x)\notin L$, for any proof $\tilde{\pi}$, we have $\Pr\left(\forall i\in V(G):V_i^{\pi}(x_{N(i)}=1\right)<s$.
\end{itemize}
for some $(V_i)_{i\in V(G)}$ with each verifier $V_i$ using at most $r$ random bits and making at most $q$ queries to $\pi$.
\end{definition}

\subsection{Constant-query $\dPCP$}
We will demonstrate constant-query $\dPCP$ protocols for the problems $\Nonbipartite$, $\Leader$, and $\Span$. As seen above, there exist no locally checkable proofs of size $o(\log n)$ for these problems, suggesting that probabilistic checking is more powerful than the fixed proofs of $\LCP$. However, the protocols below require that $G\in\mathcal{F}_n$, where $\mathcal{F}_n$ is the family of connected graphs with $n$ vertices. This is required only for the verifier to know how to query a Hadamard encoding of an $n$-dimensional vector, and it is not clear that this is central to the protocol itself.

\subsubsection{Nonbipartite}
Recall that a graph $G$ is bipartite if and only if it has no cycles with an odd number of vertices. Thus, it suffices to provide a proof that some odd cycle exists in $G$. Given some odd cycle $C$, consider the vector $\alpha_C\in\{0,1\}^n$ with $n=|V(G)|$, where $\alpha_{Ci}=1$ if and only if $i\in C$. Since we want to verify that $C$ is in fact an odd cycle in a constant number of queries, we let the proof be $\pi=\Had(\alpha_C)$ the Hadamard encoding of $\alpha_C$. Let $e_i$ be the basis vector with value $1$ at coordinate $i$ and $0$ everywhere else, and let $\mathbbm{1}$ be the all ones vector. Each verifier $V_i$ then runs the following protocol:

\begin{enumerate}
    \item Run linearity check on $\pi$.
    \item Query $a_i=\alpha_C\trans e_i$.
    \item If $a_i=1$, communicate with neighbors to ensure there exist exactly two distinct $j\in N(i)$ such that $a_j=1$.
    \item Query and check $\alpha_C\trans\mathbbm{1}=1$.
\end{enumerate}

Note that each query is error-corrected --- that is, instead of querying e.g. $\alpha\trans v$, we instead query $\alpha\trans (v+r) + \alpha\trans r$ for a randomly sampled $r\in\{0,1\}^n$. For any graph $G\in\Nonbipartite$, the prover can find some loop with odd vertices, so completeness holds. Soundness follows from the converse of this, and the fact that the linearity check and each query introduce only a constant probability of error. Therefore, $\Nonbipartite\in\dPCP_{1,\sfrac{1}{2}}[O(2^n), O(n), O(1)]$.

\subsubsection{Unique Leader}
In the constant-query protocol for $\Leader$, the honest proof is simply the Hadamard encoding of the input: $\pi=\Had(\alpha_x)$, where $\alpha_{xi}=x(i)$ for $i\in V(G)$. The protocol for each $V_i$ is as follows:
\begin{enumerate}
    \item Run linearity check on $\pi$.
    \item Query and check $\alpha_x\trans e_i=x(i)$.
    \item 
    If $x(i)=1$:
    \begin{itemize}
         \item Sample $r_i$ which is zero on entry $i$ and uniform in $\{0,1\}$ elsewhere.
        \item Query and check $\alpha_x\trans r_i=0$.
    \end{itemize}
    Else:
    \begin{itemize}
        \item Query and check $\alpha_x\trans\mathbbm{1}=1$.
    \end{itemize}
\end{enumerate}
If the prover attempts to lie about $x(i)$, it will be caught in the second step of the protocol. Otherwise, the third step verifies that there exists a unique $i$ such that $x(i)=1$ and $\forall j\neq i: x(j)=0$. Completeness and soundness follow.

\subsubsection{Spanning Tree}
Here, an honest prover sends $\pi=(\Had(\alpha_r), (\Had(\alpha_i))_{i\in V(G})$ where
\begin{equation*}
    \alpha_{ri}=\begin{cases}
     1 & \text{if $x(i)=\text{root}$}\\
     0 & \text{otherwise}
    \end{cases}
\end{equation*}
and 
\begin{equation*}
    \alpha_{ij}=\begin{cases}
    1 & \text{if $j$ reachable from $i$ in $T(x)$}\\
    0 & \text{otherwise}
    \end{cases}
\end{equation*}
where $T(x)$ is the directed graph defined on $V(G)$ by $x$, where $(i,j)\in E(T(x))$ if $x(i)=j$. The verifier runs the $\Leader$ protocol on $\Had(\alpha_r)$, then, if $x(i)$ is not root, checks $(\alpha_i+\alpha_{x(i)})=e_i$ using queries
\begin{enumerate}
    \item Check $(\alpha_i+\alpha_{x(i)})\trans e_i=\alpha_i\trans e_i+\alpha_{x(i)}\trans e_i=1$.
    \item Check $(\alpha_i+\alpha_{x(i)})\trans r_i=0$ for $r_i=0$ on element $i$ and uniform from $\{0,1\}$ elsewhere.
\end{enumerate}
It follows that $\Span\in\dPCP_{1,\sfrac{1}{2}}[2^{O(n)}, O(n), O(1)]$

\section{Conclusions \& Future Work}
We have introduced the $\dPCP$ model, which we believe to be an interesting and meaningful notion of distributed proof. Future directions include the relationship between $\dPCP$ and the $\logLCP$ or $\dIP$ classes --- although we presented some protocols for graph languages in $\logLCP$ and $\dIP$, we have yet to devise any general reductions. It could also be interesting to explore the creation of succinct arguments from $\dPCP$ using cryptographic methods, which may be useful in practical settings.


\begin{thebibliography}{}


\bibitem{lcp}
G{\"o}{\"o}s, Mika, and Jukka Suomela. ``Locally checkable proofs in distributed computing." \textit{Theory of Computing} 12.1 (2016): 1-33.

\bibitem{ram_verifier}
Naor, Moni, Merav Parter, and Eylon Yogev. ``The Power of Distributed Verifiers in Interactive Proofs." \textit{arXiv preprint arXiv:1812.10917} (2018).

\bibitem{dip_proposal}
Kol, Gillat, Rotem Oshman, and Raghuvansh R. Saxena. ``Interactive distributed proofs." \textit{Proceedings of the 2018 ACM Symposium on Principles of Distributed Computing.} ACM, 2018.

\end{thebibliography}
\end{document}